# Skyrmions at room temperature: From magnetic thin films to magnetic multilayers

C. Moreau-Luchaire[1], C. Moutafis[2], N. Reyren[1], J. Sampaio[1], N. Van Horne[1], C. A. F. Vaz[2], K. Bouzehouane[1], K. Garcia[1], C. Deranlot[1], P. Warnicke[2], P. Wohlhüter[2], J.-M. George[1], M. Weigand[3], J. Raabe[2], V. Cros[1], A. Fert[1]

[1] *Unité Mixte de Physique CNRS/Thales and Université Paris Sud, Palaiseau, France.*
[2] *Swiss Light Source, Paul Scherrer Institute, Villigen, Switzerland.*
[3] *Max Planck Institute for Intelligent Systems, Stuttgart, Germany.*

**Abstract**: Facing the ever-growing demand for data storage will most probably require a new paradigm. Magnetic skyrmions are anticipated to solve this issue as they are arguably the smallest spin textures in magnetic thin films in nature. We designed cobalt-based multilayered thin films where the cobalt layer is sandwiched between two heavy metals providing additive interfacial Dzyaloshinskii-Moriya interactions, which reach about 2 mJ/m$^2$ in the case of the Ir|Co|Pt multilayers. Using a magnetization-sensitive scanning x-ray transmission microscopy technique, we imaged magnetic bubble-like domains in these multilayers. The study of their behavior in magnetic field allows us to conclude that they are actually magnetic skyrmions stabilized by the Dzyaloshinskii-Moriya interaction. This discovery of stable skyrmions at room temperature in a technologically relevant material opens the way for device applications in a near future.

A major societal challenge is related to the continually increasing quantity of information to process and store. The hard disk drives, in which information is encoded magnetically, allow nowadays the storage of zettabytes ($10^{21}$) of information, but this technology should soon reach its limits. An up-and-coming avenue has been opened by the discovery of magnetic skyrmions [*1*], *i.e.* spin windings that can be localized within a diameter of a few nanometers and can move like particles [*2*]. These magnetic solitons, remarkably robust against defects due to the topology of their magnetic texture [*3*], are promising for being the ultimate magnetic bits to carry and store information. The topology of the skyrmions also appears to further underlie other important features such as their current-induced motion induced by small dc currents that is crucial for real applications but also the existence of a specific component in Hall Effect [*4-6*] that can be used advantageously for an electrical read-out of the information carried by nano-scale skyrmions. We proposed recently that these skyrmions could be used in future storage devices and information processing [*2*].

The existence of skyrmion spin configuration has been predicted theoretically about thirty years ago [*1*] but it was only recently that skyrmion lattices have been observed in crystals with non-centrosymmetric lattices, *e.g.* B20 crystallographic structure in MnSi [*7-9*], FeCoSi [*10*] or FeGe [*5*] crystals. In 2011, skyrmions have also been identified in single ultrathin ferromagnetic films with out-of-plane magnetization (Fe and FePd) deposited on a heavy metal substrate such as Ir(1 1 1) [*11, 12*]. Thin magnetic films appear to be more compatible with technological developments, though the observation of skyrmions in thin films has been limited up to now to low temperature and also needs, in some cases, the presence of a large applied magnetic field [*12*]. The study of these new magnetic phases associated with chiral interactions has generated a very strong interest in the community of solid state physics and magnetic systems with reduced dimensions.

The magnetic skyrmions are magnetic solitons that are, in most cases, induced by chiral interactions of the Dzyaloshinskii-Moriya (DM) type, which result from spin-orbit effects in the absence of inversion symmetry. The DM Interaction (DMI) between two neighboring atomic spins $S_1$ and $S_2$ can be expressed as: $H_{DMI} = \boldsymbol{D}_{12} \cdot (\boldsymbol{S}_1 \times \boldsymbol{S}_2)$ where $\boldsymbol{D}_{12}$ is the DM vector (Fig. 1A). In this work, we will focus on ultrathin magnetic systems in which the DMI results from the breaking of the inversion symmetry at the interface of a magnetic layer and can be large at the interface with a heavy metal having a large spin-orbit coupling [*11*]. In such case, the DM vector $\boldsymbol{D}_{12}$ is perpendicular to the $\boldsymbol{r}(S_1) - \boldsymbol{r}(S_2)$ line ($\boldsymbol{r}(\cdot)$ being the position vector) and gives rise to cycloidal skyrmionic configurations, which have a given chirality, also called hedgehog skyrmion (see Fig. 1B). Our main goal is to extend the already-observed generation of interface-induced skyrmions from single magnetic films [*11*, *12*] to multilayers by stacking layers of magnetic metals and nonmagnetic heavy metals (*e.g.* Pt and Ir) so as to induce DMI at all the magnetic interfaces (Fig.1A). The advantages of such innovative multilayered systems are twofold. Firstly we anticipate that thermal stability of skyrmions can be greatly improved, simply because of the increase of their magnetic volume, as it turns out for our samples that the same magnetic texture extends vertically throughout the multilayer. Secondly, the choice of two different heavy metals A and B sandwiching each magnetic layer potentially allows tuning the amplitude of interfacial chiral interaction, notably to increase it drastically if the two heavy materials induce interfacial chiral interactions of opposite symmetries and parallel $\boldsymbol{D}$ when they are on opposite sides of the magnetic layer. As we will see, our multilayers present bubble-like local configurations characterized by a reversed magnetization in the center [*15*].

The first important conclusion that can be draught is that the presence of these bubble-like configurations cannot be accounted for by dipolar interaction but only by the large DMI of our asymmetric multilayered systems. We will then show that the size of the bubble-like configurations and its field dependence is consistent with micromagnetic simulations of skyrmionic states induced by large DMI. This stabilization in multilayered films and at room temperature of individual magnetic skyrmions induced by chiral interactions represents the most important result of this work. Given the important features of magnetic skyrmions associated to their topological nature compared to other magnetic configurations (size, easier current-induced propagation, smaller sensitivity to defects, ...), these advances represent a definite breakthrough in the research on single skyrmion towards the potential development of skyrmion-based devices.

**Multilayers with additive chiral interaction at interfaces with heavy metal layers**

The prototype of the multilayered systems in which we aim to investigate the magnetic configuration is presented in Fig. 1A. The samples grown by sputtering deposition are stacks of 10 Ir|Co|Pt trilayers, each trilayer being composed of a 0.6 nm thick Co layer sandwiched between 1 nm of Ir and 1 nm of Pt: Pt10|Co0.6|Pt1|{Ir1|Co0.6|Pt1}$_{10}$|Pt3. The choice of the two heavy materials *i.e.* Pt and Ir, has been guided by recent experiments of asymmetric domain wall propagation [*16*, *17*] and recent *ab initio* predictions of opposite DMI for Co on Ir and Co on Pt [*18*], which corresponds to additive DMI at the two interfaces of the Co layers sandwiched between Ir and Pt. In addition to these Ir|Co|Pt asymmetric multilayers, we also prepared reference samples of Pt|Co|Pt with symmetric interfaces, in which one can expect a cancellation,

at least partial, of the interfacial DMI [*19*]. Details about the growth conditions and the characterization of their magnetic properties are presented in the supplementary materials [*14*].

From SQUID measurements on our multilayers, we deduce a magnetization at saturation of $0.96 \pm 0.10$ MA/m ($1.6 \pm 0.2$ MA/m) and an effective anisotropy of $0.17 \pm 0.04$ MJ/m$^3$ ($0.25 \pm 0.07$ MJ/m$^3$) for the Ir|Co|Pt (Pt|Co|Pt) system. These magnetic parameters have been used for *micromagnetic* simulations of their magnetic configuration [*20, 21*]. We present here simulations for the case of Co layers completely decoupled. The case taking into account the proximity-induced moment [*22, 23*] in Pt and Ir is detailed in the supplementary materials and lead to even larger estimation of the DMI amplitude [*14*].

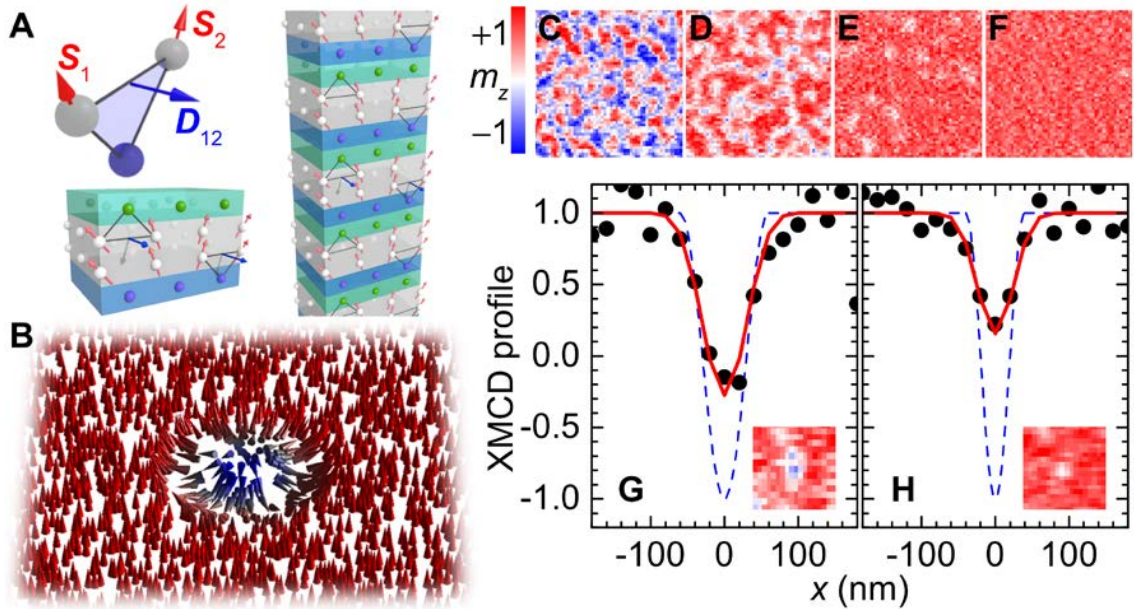

*Fig. 1*. *Interfacial Dzyaloshinskii-Moriya interaction (DMI) in asymmetric magnetic multilayers. (**A**) The DMI for two magnetic atoms close to an atom with large spin-orbit coupling in the Fert-Levy picture [13]. Zoom on a single trilayer composed of a magnetic layer (gray) sandwiched between two different heavy metals A (blue) and B (green) that induce the same chirality (same orientation of **D**) when A is below and B above the magnetic layer, and finally on an asymmetric multilayer made of several repetition of the trilayer. (**B**) Sketch of an isolated hedgehog skyrmion stabilized by interfacial chiral interaction in a magnetic thin film. (**C-F**) 800×800 nm$^2$ out-of-plane magnetization (m$_z$) map obtained by Scanning Transmission X-ray Microscopy (STXM) on a {Ir|Co|Pt}$_{10}$ multilayer at room temperature for applied out-of-plane magnetic fields of 8 (C), 38 (D), 73 (E) and 83 mT (F). (**G**) Experimental dichroic signal through a magnetic circular domain (skyrmion) as observed at 22 mT (black dots). The corresponding STXM 360×360 nm$^2$ image is in inset. The blue dashed curve is the magnetization profile of an ideal 30 nm-radius skyrmion and the red curve derives from the model described in the text. (**H**) Same type of data at 58 mT and corresponding simulation of 20 nm-radius skyrmion. The dichroic signal is not reversed due to the limited resolution of the STXM [14].*

**Magnetization mapping in asymetric multilayers: Bubbles or skyrmions?**

In order to map the distribution of the vertical component of the magnetization in our Ir|Co|Pt multilayers and to *follow* its evolution as a function of the external perpendicular field, we have performed Scanning Transmission X-ray Microscopy (STXM) experiments on samples grown on $Si_3N_4$ membranes by measuring the dichroic signal at Co $L_3$-edge [14]. In Fig. 1C-F, we display a series of images obtained at different perpendicular field values. After saturation at large negative field, we observe a domain configuration (Fig. 1C) at low positive field that combines some worm-shape domains together with other domains having almost a circular shape. Note that the magnetic contrast that we detect by STXM is an averaged value of the magnetic configuration throughout the entire multilayer. When the field is increased to $\mu_0 H_\perp = 38$ mT, the magnetic domains favored by the field extend (Fig. 1D). Before reaching the complete saturation (at about $\mu_0 H_\perp = 83$ mT, Fig.1F), isolated magnetic bubble-like domains still exist in an almost totally polarized sample as shown at $\mu_0 H_\perp = 73$ mT (Fig. 1E). We then focus on the dimension of these bubble domains as well as on their increase when the applied perpendicular field is decreased from $\mu_0 H_\perp = 73$ mT and we will show that both are consistent with a given DMI value and the corresponding skyrmion modeling. As shown in Fig. 1G and 1H, the actual diameter of the circular-shaped bubble is estimated by a fit between the experimental STXM data with the convolution of a modeled magnetization profile with the Gaussian profile of the x-ray beam [14]. We can hence precisely determine the size evolution of these isolated domains at decreasing fields from the analysis of the STXM images. Results for the Ir|Co|Pt multilayers are presented in Fig. 2. We find that the bubble diameter goes from about 30 nm (that is close to the STXM resolution) at $\mu_0 H_\perp = 73$ mT to 80 nm at $\mu_0 H_\perp = 12$ mT. For even lower field (closer to zero), the magnetic contrast evolves towards worm-like domains from which a proper diameter can no longer be defined. We emphasize that the bubble diameter (around 80 nm) at very low field value remains extremely small compared to usual values (at least larger than 800 nm with our films parameters) observed in classical bubble systems [15] in which the driving mechanism for bubble stabilization is the dipolar interaction.

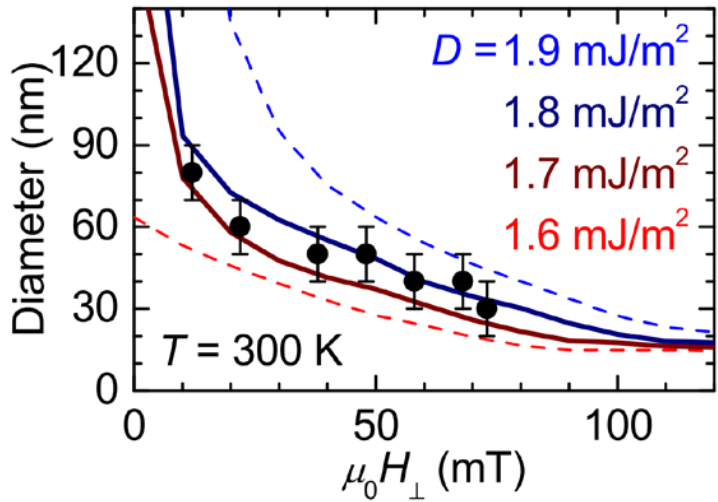

*Fig. 2.* Skyrmion diameter as a function of the external out-of-plane magnetic field $H_\perp$. Experimental diameters (dots) can be compared to micromagnetic simulations for different values of D (lines). Simulations were obtained in the case which gives a lower bound to experimental D value: totally decoupled layers.

To get more insights about the actual interaction at play in our multilayered Ir|Co|Pt systems, we compare in Fig. 2 the experimental data for the evolution of bubble size as a function of the perpendicular field with the prediction of the micromagnetic simulations of DMI-induced

skyrmions [14]. It is first important to emphasize that in these simulations, using as input the experimental parameters for the magnetization and anisotropy constant, we were not able to stabilize a single magnetic bubble without introducing a non-zero DMI. With DMI taken into account and for $D > 1.6$ mJ/$m^2$, bubble-like configurations relax and stabilize as isolated skyrmions, i.e. having a winding number $W$ equals to 1, where $W = \frac{1}{4\pi}\int \mathbf{s}\cdot(\partial_x\mathbf{s}\times\partial_y\mathbf{s})\mathrm{d}x\mathrm{d}y$. We plot in Fig. 2 simulated diameters vs $H_\perp$ curves corresponding to different values of DMI ranging from 1.6 to 1.9 mJ/$m^2$. We find that the best agreement with the experimental size and its variation with field is found for $D = 1.75 \pm 0.05$ mJ/$m^2$. This value represents one of the largest DM amplitudes ever demonstrated in chiral spin textures and is comparable for example, with what has been derived in Pt|Co|AlO$_x$ trilayers from asymmetric domain nucleation experiments [24]. Note that an even larger $D$ value is found if a coupling between the layers is introduced [14]. We also emphasize that the DM value we find is in good agreement with recent theoretical predictions if we add the DMI values predicted for Pt|Co (3 monolayers) and Ir|Co (3 monolayers) [18]. Having found such a large interfacial chiral interaction validates our approach of using asymmetric multilayers. Moreover this range of DMI value is indeed close to be optimal for the stabilization of isolated skyrmions (and so of isolated magnetic bits) instead of skyrmion arrays, as it corresponds to about 80% of the threshold value of 2.1 mJ/$m^2$ needed to obtain a skyrmion lattice ground state [25, 26].

In parallel to the measurements on asymmetric multilayers, we have performed a similar analysis with STXM images obtained on the symmetric Pt|Co|Pt systems [14]. The best fit for the size and field dependence of bubble-like configurations is obtained with DMI that are smaller but not perfectly zero, as it *would* be expected from the compensation of chiral interaction at the top and bottom Co interfaces. Indeed, in real samples, the fine (crystallographic) structure of the two interfaces might be different resulting in a finite chiral interaction as also found by other groups [17, 19].

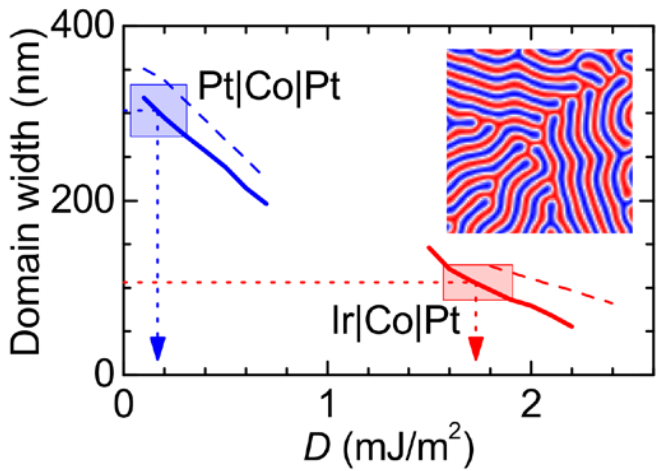

*Fig. 3. Micromagnetic simulations and experimental measurements of mean domain width evolution with DMI after demagnetization. Comparing the simulations (thick lines for the decoupled case, dashed line for the coupled case) with the experimental domain width value (dotted horizontal line) allows to estimate: D(Ir|Co|Pt) = 1.8 ± 0.2 mJ/$m^2$ and D(Pt|Co|Pt) = 0.2 ± 0.2 mJ/$m^2$. The box height represents the error margins on the experimental domain size evaluation; its width gives the error on D for the used simulation parameters. The inset shows a typical simulated worm pattern for D = 2.1 mJ/$m^2$ in Ir|Co|Pt (1.5×1.5 μ$m^2$).*

As means to confirm the large DM values in Ir|Co|Pt multilayers, we have developed a second approach to estimate them from the same experimental STXM images. This method relies on the quantitative analysis of the mean width of perpendicular magnetic domains obtained after demagnetization. From the images such as the one presented in Fig. 1C, we can estimate the

average domain width to be $106 \pm 20$ nm for the Ir|Co|Pt multilayers. The same analysis of similar images for the symmetric Pt|Co|Pt multilayers leads to a mean domain width of $303 \pm 30$ nm. In Fig. 3, these experimental values are compared with those obtained in a series of micromagnetic simulations of domains in a $2\times2$ µm$^2$ as a function of the DMI, allowing the value of *D* to be estimated [*14*]. The main result of this second approach is that the DMI is found to be about $1.8 \pm 0.2$ mJ/m$^2$ for the asymmetric Ir|Co|Pt multilayers and $0.2 \pm 0.2$ mJ/m$^2$ for the symmetric Pt|Co|Pt sample. These values obtained from domain sizes are in good agreement with those derived from the field dependence of the bubble size. As additional confirmation, we have also performed a series of magnetic images by Magnetic Force Microscopy on the very same multilayers grown by sputtering deposition on SiO$_2$/Si substrate rather than on Si$_3$N$_4$ membranes. Importantly, we find similar values of the mean domain width for images recorded at zero field after a demagnetization process, meaning that the estimated DM amplitudes are equal to those determined by STXM.

We have also evaluated how our estimation of DMI is changed if we do take into account the proximity-induced magnetization of Ir and Pt, resulting in changes of the magnetization and anisotropy parameters introduced in the simulations of the magnetic texture in the Co layers. We find that, in this case which is probably closer to what is actually happening in the multilayers, the DM amplitudes from the asymmetric multilayers is even larger and reaches about $2.1 \pm 0.2$ mJ/m$^2$. In consequence, all these results converge to the existence of very large interfacial DMI, and thus that we obtained skyrmions at room temperature in technologically relevant systems of magnetic multilayers.

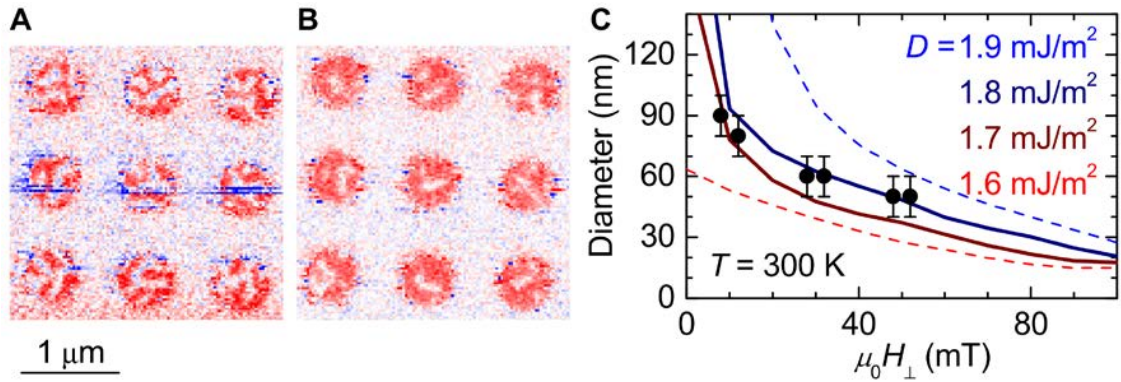

*Fig. 4. Evolution of the skyrmion size in patterned nanoscale disks. (**A**) Room temperature out-of-plane magnetization map of an array of 500 nm-diameter disk with an out-of-plane external field $\mu_0 H_\perp = 12$ mT using the same color code as in Fig.1. (**B**) Same area at $\mu_0 H_\perp = 48$ mT. (**C**) Magnetic field evolution of the skyrmion size derived from micromagnetic simulations (lines for different D values) and sizes of the observed skyrmions (symbols) for the 500 nm-diameter disks of panels A and B.*

**Room temperature observation of isolated skyrmion in magnetic disks**

In the previous section, we have deduced from two independent analyses of the magnetic configurations that very large DMI and skyrmionic structures are obtained in asymmetric {Ir|Co|Pt}$_{10}$ multilayered films. Here we demonstrate that, after patterning multilayers by e-beam lithography and ion etching, we can also stabilize isolated skyrmionic textures at room temperature in nanodisks. In Fig. 4(A-B), we first show an STXM image obtained on arrays of

500 nm-diameter disks. The region in white (zero out-of-plane magnetization) between the disks corresponds to the $Si_3N_4$ membrane. These images have been recorded after application of a large positive field. We clearly observe isolated magnetic skyrmions in most of the disks. Their typical diameter at low field is about 90 nm (Fig. 4C). From our previous numerical study of the stabilization of isolated skyrmions in magnetic nanostructures [26], we know that, for values of DMI lower than the threshold value leading to a skyrmion ground state, the final skyrmion diameter at zero field is not strongly dependent on the disk diameter for such large diameters, but rather depends on the effective ratio between DM and exchange interactions. This is consistent with the similar skyrmion diameters of the order of 80 nm found in 300 and 200 nm-diameter disks [14]. The next step has been to vary $H_\perp$ and look at the change of the skyrmion diameter, as shown in Fig. 4C for disks of 500 nm [25-27]. This field dependence of the diameter is compared to what is obtained in micromagnetic simulations performed with material parameters from experiment. Again a major outcome of these simulations is that it is not possible to stabilize any non-uniform magnetic configuration in such submicronic disks without introducing large DM values of the order of those extracted from experiments in the previous section. Moreover, the simulations show that the winding number of these circular domain is, after stabilization, always equal to one, allowing us to conclude that the experimentally observed non-collinear spin texture are magnetic skyrmions with a chirality given by *D*.

**Conclusion**

Our room temperature observation of isolated skyrmionic spin textures that are stabilized in magnetic multilayers by a strong interfacial chiral interaction represents the main achievement of this work. In turn, we hope that this experimental breakthrough can be the starting point of further fundamental studies on the very rich physics of skyrmions, as well as the development of skyrmion-based devices for memory and/or logic applications.

**Supplementary Materials:**

*Magnetic properties of Ir/Co/Pt and Pt/Co/Pt multilayers*

The studied samples are Co-based multilayers, which were grown by dc sputtering at room temperature under an Ar pressure of about 0.25 Pa on two different substrates: thermally oxidized Si wafer or 200 nm-thick $Si_3N_4$ membranes. The samples grown on $SiO_2$ were used to determine the macroscopic magnetic properties of the samples, while the ones on membranes were used for scanning transmission x-ray microscopy. A 10 nm-thick buffer made of textured Pt ((1 1 1)-oriented *fcc*) is first deposited, immediately followed by the growth of a Co(0.6 nm)|Pt(1 nm) bilayer. This first bilayer is used to induce a strong enough perpendicular magnetization anisotropy (PMA). We subsequently grew asymmetric structures {Ir(1 nm)|Co(0.6 nm)|Pt(1 nm)}$_{10}$ or symmetric ones {Co(0.6 nm)|Pt(1 nm)}$_{10}$. Both types of magnetic multilayers are finally capped by an extra 3 nm-thick Pt layer to prevent oxidation. The magnetic properties of the multilayers grown on $SiO_2$ were measured by SQUID or alternating gradient field magnetometry. The saturation magnetization and the effective anisotropy are determined from the magnetization loops with out-of-plane and in-plane magnetic fields. The configuration of their vertical magnetic domains has been investigated by Magnetic Force Microscopy (MFM).

*Description of the magnetic imaging technique*

The scanning transmission x-ray experiments have been performed on two different beamlines: X07DA (PolLux) beamline at the Swiss Light Source, Paul Scherrer Institute, Villigen, Switzerland and the Maxymus beamline BESSY II, Adlershof, Germany. The images were recorded by scanning the multilayers grown on $Si_3N_4$ membranes with an x-ray beam focused by a Fresnel zone plate, providing a resolution down to 30 nm. Circular polarized light with normal incidence is used to map the out-of-plane magnetization at the nano-scale based on the X-ray Magnetic Circular Dichroism (XMCD) effect (see Fig. S1). For imaging, we scan the samples at the Co $L_3$-edge (corresponding to the $2p$-$3d$ transitions) at 779.95 eV. In order to get some quantitative information about the DMI, we analyze the evolution of skyrmion dimensions with $H_\perp$. Assuming that the intensity of the circularly polarized x-ray beam going through a layer is given by $I = I_0 \exp(-\mu t)$, where $I_0$ is the intensity of the incoming light, $\mu$ the absorption and $t$ the thickness of the layer, it can be demonstrated that the quantity $\mu_p - \mu_m$ (index $m$ and $p$ for the different helicities) is proportional to the out-of-plane magnetization $m_z$. This remains true even in the case of an x-ray beam that is not 100% circularly polarized or if the positive helicity has not the same polarization level than the negative one (because we also scanned areas of membrane without magnetic material for the normalization). Therefore, all the experimental maps of the magnetic configurations (see for example Fig. 1D-F) presented in this work are obtained by calculating the maps of the absorption difference $\mu_p - \mu_m$. Imaging of smaller magnetic disks is presented in Fig. S2.

*Determination of skyrmion diameter from STXM images*

The actual diameter of the observed magnetic domains are found to be often smaller than the actual dimension of the polarized x-ray beam (in the presented STXM experiments, the full width at half maximum (FWHM) of the beam was either 45 or 91 nm). However assuming a typical circular shape of the magnetic skyrmions and convoluting it with a Gaussian x-ray beam profile, we can fit our observations and thus deduce the actual skyrmion diameters down to 20 nm (see Fig. S3). More precisely, the assumed magnetization profiles are corresponding to approximated circular skyrmions for which the magnetization m is rotating continuously with the distance from skyrmion center $r$, *i.e.* $m_z(r) = \cos(\pi r/(2r_S))$ for $r < 2r_S$ [28]. To improve the signal/noise ratio, we average over a few lines, typically corresponding to the beam FWHM. The skyrmion profiles are then compared to the ones determined in micromagnetic simulations including DMI. In Fig. S3, we illustrate such image analysis and comparison to micromagnetic simulations for two different perpendicular applied magnetic fields for the Pt|Co|Pt multilayers.

*Quantitative estimation of DMI and role of proximity induced moment of the heavy atoms*

As presented in the main text, we have used two different approaches to extract the DM amplitude from the analysis of the STXM images *i.e.* the evolution of skyrmion diameter with $H_\perp$ (see Fig. 2), and the evolution of domain width with the DMI (see Fig. 3). Indeed, the quantitative estimation of the DMI is obtained by comparison with micromagnetic simulations that have been performed using two different solvers, the Object-Oriented MicroMagnetic Framework (OOMMF) [*20*] version 1.2a4 and MuMax3 [*21*] version 3.5 and 3.6.1, which are both taking into account the interfacial DM energy. In order to perform these simulations, we have to introduce input material parameters that are the saturation magnetization $M_s$ and the magnetic anisotropy $K$. The other parameters are the exchange constant that we take to be equal to $A = 10$ pJ/m and the magnetic damping $\alpha = 0.3$. For practical purpose in the micromagnetic simulation of skyrmion in perpendicular magnetic field, we relaxed the state at each magnetic field step by relaxing an initial state, which is already a skyrmion. We checked that, with our material properties, trivial bubbles turn to skyrmions or vanish under high magnetic field ($\approx 50$ mT or more). Using our experimental determination of the saturation, magnetization and the perpendicular anisotropy of the multilayers, two extreme hypotheses have been considered. In a first series of simulations, we suppose that the measured magnetization is confined exclusively in the Co layers. Consistently the effective magnetization of Co is simply obtained by dividing the total magnetization by the total Co thickness, *i.e.* 6.6 nm. The effective anisotropy constant is deduced from the saturation field measured in the in-plane magnetic hysteresis curve. As presented in the main text and reproduced in Fig. S4, the best agreement between experimental data and simulations is obtained for $D = 1.8 \pm 0.2$ mJ/m². However it is well known that both Pt and Ir atoms can acquire a small magnetic moment when they are in contact with a magnetic material [*22, 23*]. In such case, it results that the Co magnetization is overestimated. Then a second type of micromagnetic simulations of the magnetic texture was performed in which the proximity-induced moment in Pt and Ir is taken into account. In this case, the effective magnetization (and the effective anisotropy thereof) is obtained by dividing the total magnetization by the total multilayer thickness, *i.e.* 28.6 nm. As shown in Fig. S4, it leads to an even larger estimation of the DM amplitude from the evolution of the skyrmion diameter with field with a value equal to $D = 2.2$ mJ/m² for the Ir|Co|Pt multilayer.

**Supplementary Figures:**

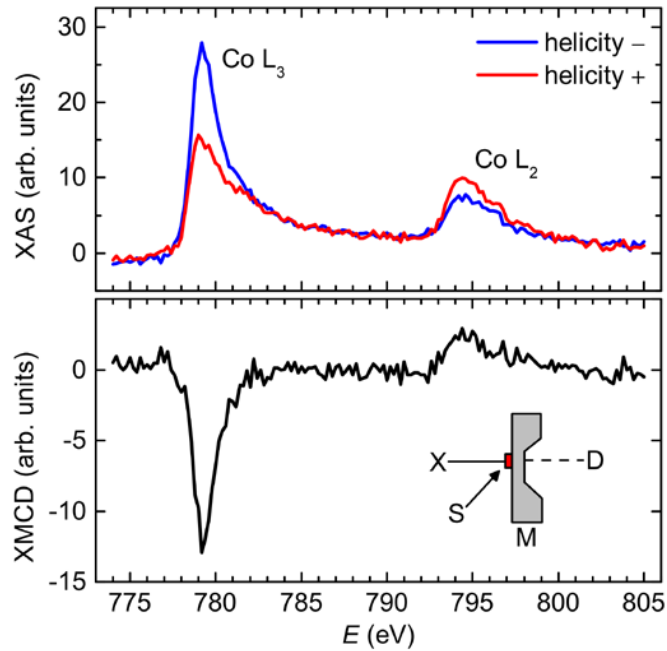

***Fig. S1.*** *XAS of the $L_{3,2}$ peaks of Co and the corresponding XMCD obtained on $\{Co0.6/Pt1\}_{10}$ at the PolLux beamline. The XAS were obtained by normalizing the intensity by the intensity of the nearby membrane and then offset to bring the pre-edge value to zero. XMCD is obtained from the difference of these XAS. The inset shows a very schematic cut diagram of the incoming x-ray beam (X), the sample S on top of the membrane M, and the photon counter D.*

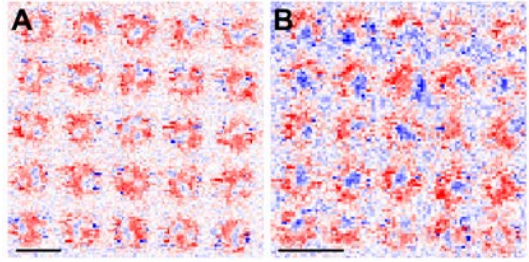

*Fig. S2.* *Image of 300 (**A**) and 200 nm-diameter (**B**) disks. The scale bars are 500 nm.*

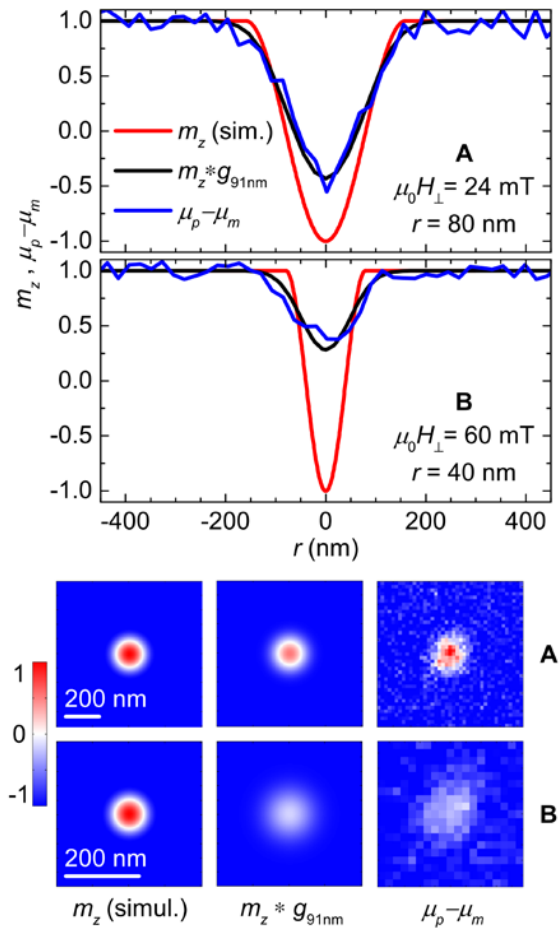

*Fig. S3.* *Actual magnetization profile of skyrmions: Cross section of a skyrmion recorded in the Pt/Co/Pt multilayer obtained for $\mu_0 H = 24mT$ (**A**) and $\mu_0 H = 60mT$ (**B**). We plot the simplified expected skyrmion magnetic profile (red curves), in black this profile convoluted with a 2D-Gaussian of FWHM of 91 nm averaged over four lines (black curves), and the actual profile also average over four lines (blue curves). In the bottom, we display the corresponding magnetization maps.*

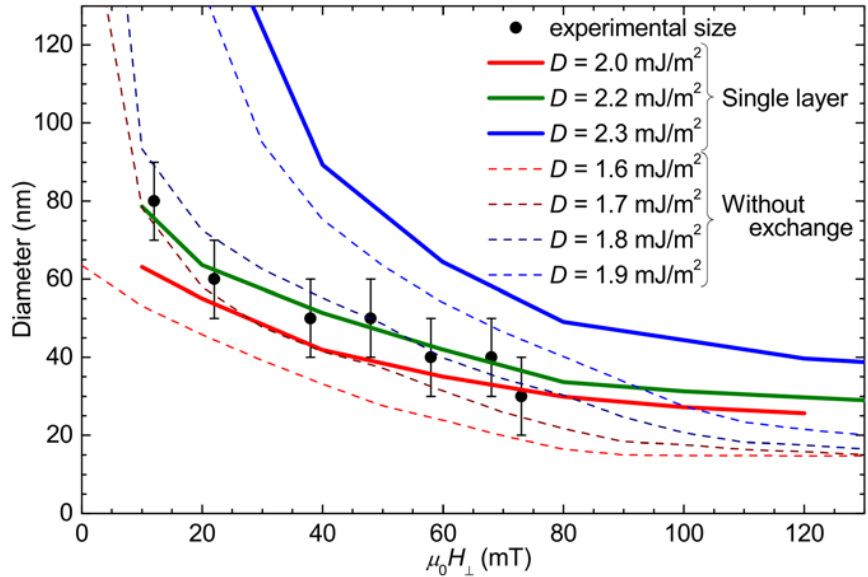

***Fig. S4.*** *Comparison of the skyrmion diameter in Ir/Co/Pt as a function of the magnetic field for different simulation parameters. (See text for details.)*

**References :**

1. A. Bogdanov, A. Yablonskii, Thermodynamically stable "vortices" in magnetically ordered crystals. *JETP Lett.* **68**, 101-103 (1989).

2. A. Fert, V. Cros, J. Sampaio, Skyrmions on the track. *Nature Nanotech.* **8**, 152-156 (2013).

3. N. Nagaosa, Y. Tokura, Topological properties and dynamics of magnetic skyrmions. *Nature Nanotech*. **8**, 899-911 (2013).

4. S. Mühlbauer, B. Binz, C. Jonietz, C. Pfleiderer, A. Rosch, A. Neubauer, R. Georgii, P. Böni, Skyrmion Lattice in a Chiral Magnet. *Science* **323**, 915-919 (2009).

5. S. X. Huang, C. L. Chien, Extended skyrmion phase in epitaxial FeGe(111) thin films. *Phys. Rev. Lett.* **108**, 267201 (2012).

6. R. Ritz, M. Halder, M. Wagner, C. Franz, A. Bauer, C. Pfleiderer, Formation of a topological non-Fermi liquid in MnSi. *Nature* **497**, 231-234 (2013).

7. A. Neubauer, C. Pfleiderer, B. Binz, A. Rosch, R. Ritz, P. G. Niklowitz, P. Böni, Topological hall effect in the a phase of MnSi. *Phys. Rev. Lett*. **102**, 186602, (2009).

8. C. Pappas, E. Lelièvre-Berna, P. Falus, P. M. Bentley, E. Moskvin, S. Grigoriev, P. Fouquet, B. Farago, Chiral paramagnetic skyrmion-like phase in MnSi. *Phys. Rev. Lett.* **102**, 197202 (2009).

9. A. Tonomura, X. Yu, K. Yanagisawa, T. Matsuda, Y. Onose, N. Kanazawa, H. S. Park, Y. Tokura, Real-space observation of skyrmion lattice in helimagnet MnSi thin samples. *Nano Letters* **12**, 1673-1677 (2012).

10. X. Z. Yu, Y. Onose, N. Kanazawa, J. H. Park, J. H. Han, Y. Matsui, N. Nagaosa, Y. Tokura, Real-space observation of a two-dimensional skyrmion crystal. *Nature* **465**, 901-904 (2010).

11. S. Heinze, K. von Bergmann, M. Menzel, J. Brede, A. Kubetzka, R. Wiesendanger, G. Bihlmayer, S. Blügel, Spontaneous atomic-scale magnetic skyrmion lattice in two dimensions. *Nature Phys.* **7**, 713-718 (2011).

12. N. Romming, C. Hanneken, M. Menzel, J. E. Bickel, B. Wolter, K. von Bergmann, A. Kubetzka, R. Wiesendanger, Writing and deleting single magnetic skyrmions. *Science* **341**, 636-639 (2013).

13. A. Fert, P. M. Levy, Role of anisotropic exchange interactions in determining the properties of spin-glasses. *Phys. Rev. Lett.* **44**, 1538-1541 (1980).

14. See Supplementary Materials

15. A. P. Malozemoff, J. C. Slonczewski, *Magnetic Domain Walls in Bubble Materials* Raymond Wolfe Ed. (Academic Press, New York, 1979).

16. Y. P. Kabanov, Y. L. Iunin, V. I. Nikitenko, A. J. Shapiro, R. D. Shull, L. Y. Zhu, C. L. Chien, In-plane field effects on the dynamics of domain walls in ultrathin co films with perpendicular anisotropy. *IEEE Trans. Magn.* **46**, 2220-2223 (2010).

17. A. Hrabec, N. A. Porter, A. Wells, M. J. Benitez, G. Burnell, S. McVitie, D. McGrouther, T. A. Moore, C. H. Marrows, Measuring and tailoring the Dzyaloshinskii-Moriya interaction in perpendicularly magnetized thinfilms. *Phys. Rev. B* **90**, 020402 (2014).

18. H. Yang, A. Thiaville, S. Rohart, A. Fert, M. Chshiev, Anatomy of Dzyaloshinskii-Moriya Interaction at Co/Pt Interfaces. arXiv:1501.05511 (2015).

19. J. H. Franken, M. Herps, H. J. M. Swagten, B. Koopmans, Tunable chiral spin texture in magnetic domainwalls. *Sci. Rep.* **4**, 5248 (2014).

20. M. J. Donahue, D. G. Porter, "Oommf user's guide version 1.0". (Interagency Report NISTIR 6376, National Institute of Standards and Technology, Gaithersburg, MD, 1999).

21. A. Vansteenkiste, J. Leliaert, M. Dvornik, M. Helsen, F. Garcia-Sanchez, B. Van Waeyenberge, The design and verification of mumax3. *AIP Advances* **4**, 107133 (2014).

22. W. Grange, M. Maret, J.-P. Kappler, J. Vogel, A. Fontaine, F. Petroff, G. Krill, A. Rogalev, J. Goulon, M. Finazzi, N. B. Brookes, Magnetocrystalline anisotropy in (111) $CoPt_3$ thin films probed by x-ray magnetic circular dichroism. *Phys. Rev. B*, **58**, 6298-6304 (1998).

23. K.-S. Ryu, S.-H. Yang, L. Thomas, S. S. P. Parkin, Chiral spin torque arising from proximity-induced magnetization. *Nature Commun.* **5**, 3910 (2014).

24. S. Pizzini, J. Vogel, S. Rohart, L. D. Buda-Prejbeanu, E. Jué, O. Boulle, I. M. Miron, C. K. Safeer, S. Auffret, G. Gaudin, A. Thiaville, Chirality-induced asymmetric magnetic nucleation in Pt/Co/AlOx ultra-thin microstructures. *Phys. Rev. Lett.* **113**, 047203 (2014).

25. S. Rohart, A. Thiaville, Skyrmion confinement in ultrathin film nanostructures in the presence of Dzyaloshinskii-Moriya interaction. *Phys. Rev. B* **88**, 184422 (2013).

26. J. Sampaio, V. Cros, S. Rohart, A. Thiaville, A. Fert, Nucleation, stability and current-induced motion of isolated magnetic skyrmions in nanostructures. *Nature Nanotech.* **8**, 839-844, (2013).
27. N. S. Kiselev, A. N. Bogdanov, R. Schäfer, U. K. Rößler, Chiral skyrmions in thin magnetic films: new objects for magnetic storage technologies? *J. Phys. D: Appl. Phys.* **44**, 392001 (2011).
28. A. N. Bogdanov, U. K. Rößler, Chiral symmetry breaking in magnetic thin films and multilayers. *Phys. Rev. Lett.* **87**, 037203 (2001).

**Acknowledgments:** The authors acknowledge technical support from Blagoj Sarafimov and Michael Bechtel for their technical support at the SLS and Bessy II beamlines. The STXM experiments were performed using the X07DA (PolLux) beamline at the Swiss Light Source, Paul Scherrer Institut, Villigen, Switzerland and the Maxymus beamline BESSY II, Adlershof, Germany. The authors acknowledge financial support from the ANR agency ANR-14-CE26-0012 (ULTRASKY).

**Author Contributions:** NR, CM, VC and AF conceived the project. CD and CML grew the films. CAV, CM, KG patterned the samples. CM, CML, NR, JS, NVH, CAV, KB, PW, PW, MW, JR and VC acquired the data at the synchrotron. CML and NR treated and analyzed the data with the help of CM and PW. CML, JS and NR performed the micromagnetic simulations. CML, NR and VC prepared the manuscript. All authors discussed and commented the manuscript.